# Systems Computing Challenges in the Internet of Things

**Rajeev Alur, Emery Berger, Ann W. Drobnis, Limor Fix, Kevin Fu, Gregory D. Hager, Daniel Lopresti, Klara Nahrstedt, Elizabeth Mynatt, Shwetak Patel, Jennifer Rexford, John A. Stankovic, and Benjamin Zorn**

**September 22, 2015**

## 1. Introduction

A recent McKinsey report estimates the economic impact of the Internet of Things (IoT) to be between $3.9 to $11 trillion dollars by 2025[1]. IoT has the potential to have a profound impact on our daily lives, including technologies for the home, for health, for transportation, and for managing our natural resources. The Internet was largely driven by information and ideas generated by *people*, but advances in sensing and hardware have enabled computers to more easily observe the *physical world*. Coupling this additional layer of information with advances in machine learning brings dramatic new capabilities including the ability to capture and process tremendous amounts of data; to predict behaviors, activities, and the future in uncanny ways; and to manipulate the physical world in response. This trend will fundamentally change how people interact with physical objects and the environment. Success in developing value-added capabilities around IoT requires a broad approach that includes expertise in sensing and hardware, machine learning, networked systems, human-computer interaction, security, and privacy. Strategies for making IoT practical and spurring its ultimate adoption also require a multifaceted approach that often transcends technology, such as with concerns over data security, privacy, public policy, and regulatory issues[2].

In this paper we argue that *existing best practices in building robust and secure systems are insufficient to address the new challenges that IoT systems will present*. We provide recommendations regarding investments in research areas that will help address inadequacies in existing systems, practices, tools, and policies. The goal of this white paper is to consider the core software, systems, and networking technology shifts created by the IoT trend and try to anticipate the major challenges such systems face in terms of usability, performance, security, and reliability. There are many important research challenges beyond those discussed here that need to be addressed for IoT to see its full potential. Specifically, topics like data management, storage and communication, machine learning, and privacy are also important but not discussed here. A recent report from the Semiconductor Industry Association and Semiconductor Research Corporation covers a number of the topics not considered here.[3]

---

[1] *The Internet of Things: Mapping the Value Beyond the Hype*, McKinsey Global Institute, June 2015, http://www.mckinsey.com/insights/business_technology/the_internet_of_things_the_value_of_digitizing_the_physical_world

[2] *Internet of Things: Privacy and Security in a Connected World*, Federal Trade Commission, Jan. 2015. https://www.ftc.gov/reports/federal-trade-commission-staff-report-november-2013-workshop-entitled-internet-things

Because IoT can mean many different things, to make the discussion concrete, we illustrate the architecture of a consumer-facing IoT system shown below.

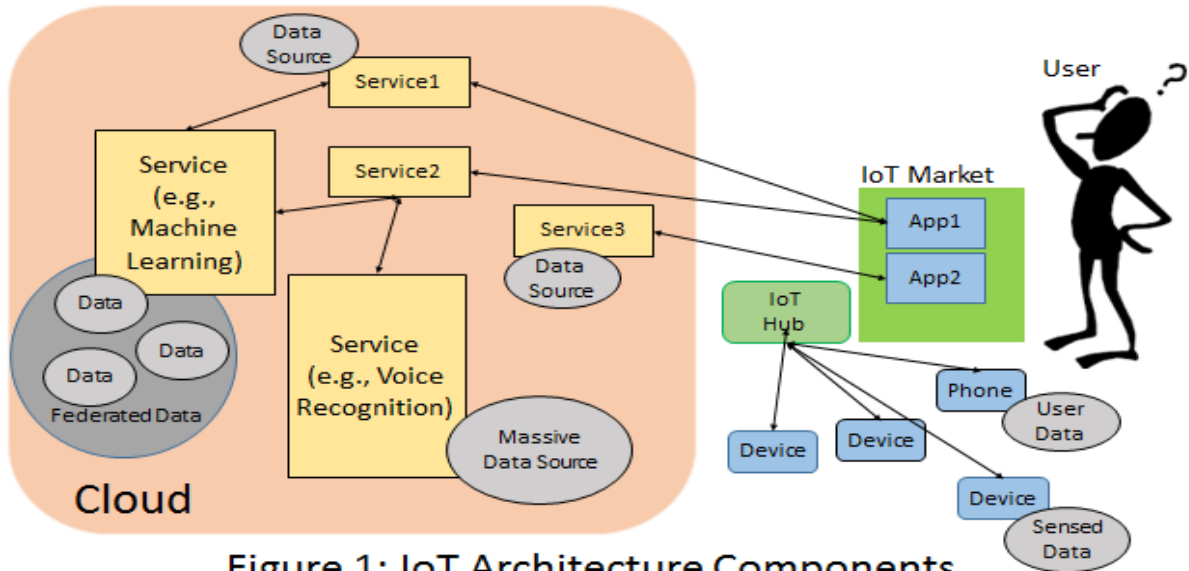

Figure 1: IoT Architecture Components

Figure 1 shows likely components of a generic IoT system and how they interact. The major components already exist in specific instances and currently companies are competing to become the de facto platform for such devices.

The components are:

- Hardware devices that are able to sense and interface with the physical world
- Data collected on the behalf of the user by these devices
- IoT hubs that funnel data from the physical world to the cloud
- An IoT marketplace with value-added apps that interact with devices and the cloud
- Services, large and small, that the apps connect to (could be one or more, could be a vertical device-app-service, or could be stratified)
- Varying sizes of data stores, including federated data stores that normalize data from heterogeneous sources, that the services maintain (collected from the apps and devices)

The rest of the document gives examples of the opportunities for IoT and then highlights important technical challenges present in these systems that may limit its potential.

---

[3] *Rebooting the IT Revolution: A Call to Action*, August 2015, Semiconductor Industry Association and Semiconductor Research Corporation. https://www.src.org/newsroom/rebooting-the-it-revolution.pdf

## 2. Opportunities

Before discussing the challenges, we consider specific scenarios that show the enormous potential of these systems.

**Opportunity: An Internet of Healthcare Things**

An Internet of Healthcare Things (IoHT) can spur a revolution in medicine, healthcare delivery and consumer health. Smart medical devices, including smartphones, watches, and other bio-based wearables connected in an IoHT, can provide improved, pervasive, cost-effective, and personalized medical care and wellness. An IoHT can also improve hospitals, nursing homes, assisted living, and continuous care retirement communities in many ways. For example, in hospitals automated hand washing systems, caregiver reminders, locating devices, and automated linking of device-produced medical data with medical records can improve the operation and safety of hospitals. Further, an IoHT can not only monitor, but also support interventions and assistance. Plug-n-play devices, usable interfaces for not only the healthy, but also for the senior adults and people with disabilities, techniques to handle large volumes of data to avoid overwhelming caregivers, and studies that demonstrate the true medical value of the IoT technology are needed for the large-scale adoption of IoT home-based healthcare. A true global IoT infrastructure has the potential to revolutionize the practice of medicine and transform how people manage their health.

**Opportunity: Smart Homes**

Smart homes have been a dream for decades, but the lack of compelling user experiences and practical technology (e.g., low cost, easy-to-deploy, low maintenance, etc) has often limited large-scale deployments and mainstream adoption. There are over a 120 million homes in the U.S. and far fewer brand new homes are going to be built in the next couple decades. Thus, retrofitting is one critical problem. Emerging IoT technologies have the possibility to address this challenge by enabling consumers to instrument their own homes. Beyond just the convenience of remote door locks and alarms, instrumented homes can provide peace of mind for homeowners on the health and condition of their home and warn of emerging problems before they become costly or catastrophic. Homeowners can also use IoT technology to monitor and manage energy and water usage in the home. As the utility grids gets "smarter," the home also has to get smarter alongside to take advantage of energy reduction and conservation strategies. The plethora of IoT solutions currently on the market has been a playground for researchers to begin to investigate and study compelling use cases in the wild. IoT technology for the home is going to be critical in informing the technologies that eventually get embedded into homes and home appliances. However numerous challenges remain as the "Internet of Way Too Many Things" overemphasizes cool and costly gadgets in contrast to simple designs that integrate cleanly into home life. (NYT Sunday, 9/4) Privacy concerns grow as IoT devices are acquired by large companies, such as Google's acquisition of Nest. Finally interoperability in the home continues to be a barrier compared to other IoT domains (health, city) that provide more opportunities for professional system support and maintenance.

**Opportunity: Smart Cities**

Smart cities provide a major use case for the Internet of Things. The previously cited McKinsey report identifies cities as the second or third largest target area for IoT, with a projected economic impact totaling somewhere between $1 trillion and $1.6 trillion by 2025. Pilots abound: in China and India alone, there are nearly 300 smart city activities[4]. Smart City Barcelona has been in existence since 2012 and currently has 83 separate projects that fit into one of twelve target areas: environmental, information and communications technology, mobility, water, energy, matter (waste), nature, built domain, public space, open government, information flows, and services[5]. Decreased costs and increased revenues from the project have had a net positive impact on Barcelona's budget measuring $100 million and have reportedly contributed nearly 50,000 new jobs. Singapore is making major investments toward becoming what they term a "Smart Nation," providing open access to a vast array of government data from many sources, much of it in real time, in a strategic effort coordinated out of the Prime Minister's Office.[6] Improved quality of life is also frequently cited as major (if hard-to-measure) benefit. The tremendous potential of the smart city movement, however, brings with it a wide range of challenges that require concerted research efforts in systems computing. Many of the advantages of smart cities accrue from our new-found ability to monitor and track the activities of citizens on a massive scale, in spaces that, while public, have traditionally enjoyed significant anonymity. This pervasive tracking of citizens raises serious issues of privacy and security. Given the very large number of systems within a city that could potentially be brought online, standards for communications protocols and data sharing will play an important role. The sheer quantities of data that can be collected throughout a modest-sized city will present technical challenges in storage and organization. Administration of city resources has always been a matter of managing tradeoffs with the goal of delivering services to citizens at cost-benefit ratios they consider acceptable: the rapid, highly disruptive nature of the transition to smart cities along with a lag in technical expertise among city leaders could lead to lost opportunities and perhaps even degradations in quality of life from the very inventions intended to improve it. Education and civic participation have important roles to play here, as do technologies for helping humans to understand the complex functioning of city systems and the data they generate, and to participate in using this data for decision-making.

3.  **Cross-cutting Technical Challenges**

While IoT systems will be deployed in different vertical domains, fundamental research challenges will cut across many of the application domains. Here we consider specifically challenges in networking, security, software development, distributed systems, and cyber-physical systems. Our intent with this overview is not to present a comprehensive list of all challenges, but to highlight a few key problems in each core area.

**Networking Challenges**

Today's networking technology was not designed to support a huge number of low-power, possibly

---

[4] "*Rethinking Smart Cities From The Ground Up*", http://www.nesta.org.uk/publications/rethinking-smart-cities-ground
[5] "*IoE-Driven Smart City Barcelona Initiative Cuts Water Bills, Boosts Parking Revenues, Creates Jobs & More*"
http://www.cisco.com/assets/global/ZA/tomorrow-starts-here/pdf/barcelona_jurisdiction_profile_za.pdf
[6] "*Smart Nation Singapore: Many Smart Ideas, One Smart Nation*"
http://www.pmo.gov.sg/smartnation

mobile, devices that interact with the physical world, human users, and the cloud in sophisticated ways. These new requirements raise several important research challenges in the area of networked systems:

**Scalability**: Computer networking researchers often grapple with questions of scale. For example, today's Internet routing system interconnects more than 3 billion people, more than a half million IP address blocks, and more than 50 thousand separately administered networks. The networking community responded to the rapid growth of the Internet by designing routing protocols, router architectures, and operational practices to manage this kind of scale. But, today's network protocols were designed for a world of largely stationary devices with reasonable computational and memory resources. IoT systems will require us to go much further, to handle orders of magnitude more devices, many of which are mobile, intermittently connected, and low power. A truly seamless Internet of Things requires future network protocols and network architectures that can rise to these scalability challenges.

**Multitenancy**: IoT would benefit from ways to let multiple applications (and sets of associated IoT devices) control their own fate over a shared network infrastructure. For example, a medical equipment company may have many devices (from patients' pacemakers to MRI machines) that rely on the hospital's computer network for communication. If the hospital IT staff misconfigures the network, these medical devices effectively don't work. Similar issues arise in the smart grid, where energy companies may rely on the customers' broadband network to communicate with smart meters and smart devices (e.g., air conditioners). We need effective ways to offer virtual network infrastructure -- much like today's cloud providers give each tenant its own abstract view of the data center -- to different apps, devices, and services. Each virtual network should have its own configuration, and guaranteed share of resources, but be mapped underneath to a shared physical infrastructure.

**Network security:** IoT raises a wide range of security challenges, as discussed in the Security Challenges section below. The unique properties of IoT devices have the potential make the underlying network an even more important part of any viable defense. IoT devices may not defend themselves appropriately, due to limited computing and power resources as well as a lack of security expertise among device manufacturers and end users. The network is the common infrastructure connecting these devices to each other, and to the rest of the Internet and the cloud that stores and analyzes data. The centrality of the network creates an opportunity for new research on anomaly detection, intrusion detection/prevention systems, and access control, so that the network can block unwanted traffic or detect suspicious behavior. While these are old topics in network security, IoT creates new opportunities because many IoT devices have a narrow purpose (e.g., a picture frame that displays photos stored by a particular cloud provider) that lead to distinctive traffic patterns, perhaps enabling different approaches to detecting anomalies.

**Open network interfaces**: Today's IoT landscape is already awash in proprietary technologies and competing standards. The history of computer networking has shown the importance of open interfaces to innovation. Where we have open interfaces and programmability, we have innovation. A case in point is end-host computers, where the open standards for Ethernet, IP, TCP, and even applications (e.g., HTTP) have led to tremendous innovation. However, innovation inside the Internet has been crippled by closed, proprietary software, until the recent efforts in Software Defined Networks (SDNs) (e.g., the OpenFlow standard for interacting with the packet-forwarding logic in network switches). The networking research community can, and should, play a lead role in designing and experimenting with open APIs and protocols for the networks that interconnect IoT devices and connect IoT devices to the Internet and the cloud.  Otherwise, the natural business incentives of IoT vendors could lead to (multiple) closed environments that stifle innovation and limit interoperability.

**Low-power communication**: Many IoT devices are small and do not have access to a continuous power source.  Battery size, lifetime, and cost impose significant constraints on how these devices compute and communicate.  Novel wireless networking solutions can address these challenges.  For example, recent work shows how to avoid using power-hungry transmitters by leveraging the backscatter of radio noise from TV stations and cell towers as both an energy source and a communication medium.  Going further, many IoT devices serve a single, limited purpose, suggesting that these devices could have customized network interfaces, operating systems, and programming models that make the most effective use of limited computation, network, and energy resources.  Research in these areas involves interdisciplinary collaboration in signal processing and wireless communication, as well as computer architecture and operating systems.

**Security Challenges**

Many of the most important uses of IoT systems have significant security requirements.  If the security of IoT systems is not improved over the current state-of-the art, many important IoT applications will not be possible.  Already, law enforcement agencies are cautioning consumers of the dangers of embracing IoT technology, including a recent FBI Public Service Announcement[7] that suggests consumers should "Isolate IoT devices on their own protected networks" and "...be aware of the capabilities of the devices…", suggestions that may be very hard for many individuals to achieve in practice.

Years of experience have led to best practices that create high assurance software in some domains, such as aircraft control software, but only at a high cost and with constraints on complexity. IoT systems will have significantly greater complexity with many low-cost components. Beyond the security challenges mentioned in other parts of this document, we identify specific additional challenges here.

**Diverse, interacting, potentially unsecure devices:**  IoT raises a wide range of serious security challenges, since many IoT devices interact closely with the physical world.  Recent news has highlighted many opportunities for attack on networked cars, power stations, and implanted medical devices. The security problem is exacerbated by the fact that many IoT devices may be  built by companies that have little expertise in security, using potentially old operating systems and libraries that are not fully patched. Furthermore, if a device relies on open-source software with vulnerabilities (e.g., the many routers that used the version of OpenSSL that was susceptible to the Heartbleed virus), updating the firmware on such devices can be difficult.

**Devices that misrepresent themselves:** Another risk lies in the potential for these diverse devices to be intentionally programmed to "cheat" as was the recent case where Volkswagen was found to have programmed their software to cheat on emissions tests[8]. By cheating, we mean any action that intentionally misrepresents the product's behavior for the purpose of deceiving regulators or consumers. Examples of such cheating might be misrepresenting network bandwidth usage or performance on benchmarks. As we cede more control to these devices the need to regulate them will increase, which will give manufacturers more temptation to cheat. While the brand risk of the exposure of cheating is very high for Volkswagen, it is much lower for IoT startups, so the temptation to cheat will be even greater. Technologies, procedures, and policies are needed to allow inexpensive and effective auditing of the software in such devices, including methods to specify the expected correct behavior and solutions that allow for inspection of the product source code.

---

[7] *Internet of Things Poses Opportunities for Cyber Crime*, Federal Bureau of Investigation Public Service Announcement, Sept. 10, 2015. http://www.ic3.gov/media/2015/150910.aspx

**Security threats from ubiquitous devices**: In a world where we are surrounded by IoT devices, the ability to limit our exposure to them decreases. If a desktop computer becomes infected, we can reboot it, run an anti-virus program, and hope the problem goes away. If one or more devices in a network of IoT devices is compromised, it may be both very difficult to know what device has been compromised or how to fix the problem to restore the overall system security (for example, you may not be able to "turn everything off" for various reasons). Consider how current ransomware, which holds our data hostage, might be transformed to an attack that requires us to pay money to enter our own house or turn on the heat. Research on systematic methods for restoring IoT systems from a known good state is needed as well as tools to isolate and correct individual compromised components within the distributed system.

**System-wide security abstractions**: Programming languages have evolved to incorporate features that increase productivity and reduce classes of errors. For example, Java and C# have features that prevent errors such as buffer overflows by construction – all valid programs are correct with respect to memory safety. Next-generation IoT systems, that involve physical interaction, need to have a new generation of system-wide properties (e.g., to guarantee physical safety) that are correct by construction and checked automatically. These properties involve major improvements in our ability to reason about the interaction between the software in the system and the physics of their real-world actions (see Cyber-Physical-Human Systems below).

**Software Development**

Developing software for complex IoT systems correctly and cheaply requires new approaches to software development including abstractions, tools, and practices that reflect the changing needs of developers.

**Understanding code+data**: Increasingly the data that the program processes has a critical effect on the result of the program. IoT systems collect data and process it continuously, often then making critical decisions based on it such as whether to apply brakes in a self-driving car or whether to allow someone to unlock a door. Data can be invalid for numerous reasons: sensors can fail, users with malicious intent can inject incorrect data, and machine learning models that are not well implemented can behave incorrectly. Understanding the impact of bad data on IoT systems requires new approaches to capturing the relation between the data and the code as well as proof methods to establish the likelihood that properties of systems hold in the presence of bad data and/or incorrect code.

---

[8] "Volkswagen Says 11 Millions Cars Worldwide Are Affected in Diesel Deception", New York Times, Sept. 22, 2015. http://www.nytimes.com/2015/09/23/business/international/volkswagen-diesel-car-scandal.html

**Correctly configuring dynamic compositions of systems**: IoT systems will include increasing numbers of devices that interact in rich and complex ways. A major challenge in enabling such systems is to provide tool support to guarantee that they are configured correctly at all times. Configurations, including the software, hardware, and networking may be in constant flux and only with automation can we have any hope that such systems will remain correctly configured. Higher-level policy languages should allow users and administrators to specify high-level intent rather than configure low-level mechanisms. Recent research in Software-Defined Networking (SDN) is making progress on these challenges, through interdisciplinary collaboration with researchers in the programming languages and formal verification communities. Further research in this area, as well as stronger connections with the HCI (Human Computer Interaction) community, are essential to make IoT devices operate seamlessly and securely.

**Debugging, self-diagnosing, and automatic repair**: IoT devices will live in a complex environment and are bound to encounter situations that were not anticipated by their designers. Debugging and updating the code running on these devices will be a necessity, but will present numerous challenges. These devices will be intermittently connected to the Internet and will have limited bandwidth, making interactive debugging difficult. They are likely to be power and memory constrained, limiting their ability to track and store detailed logs of their activity for forensic analysis. Finally, opening up these devices to remote debugging and software updates will be necessary but poses significant privacy and security challenges.

The space program has tackled the challenge of debugging at a distance and remote software updating for devices like the Mars Land Rover; however, we will need to drastically lower the cost and simplify the incorporation of these techniques into programming language and runtime platforms for IoT devices. To reduce the frequency and increase the effectiveness of debugging, IoT devices will need to constantly monitor their own "health" while permitting limited, focused telemetry for debugging purposes. They will need to be equipped with software systems that can learn from errors and automatically repair themselves. In addition to verifying the source of software updates, they should conduct experiments to verify whether the patches are successful at resolving identified problems without introducing new ones.

**New and Complex Dependencies**

Future IoT systems will combine billions of sensors and actuators in an infrastructure where millions of independently developed applications (apps) will be running, some performing safety-critical functions. These new architectures give rise to many new and complex forms of systems of systems (and apps across apps) dependencies. In the past, most systems of systems composition was performed by some team to explicitly combine these systems. This team could then attempt to address dependencies at design time. In the new IoT world many systems will be co-located and be running concurrently without the explicit design time attempt to combine them. In addition to existing problems with dependencies, such as wireless interferences and mismatched APIs, these new systems will have significant new issues including open environments, humans-in-the-loop, and a richer semantics of apps individually and in combination.

**From closed to open environments:** Traditionally, the majority of sensor and actuation systems have been closed systems. For example, in the past, cars, airplanes and ships have had networked sensor systems that operate largely within that vehicle. However, these systems' capabilities and many emerging systems that support smart homes and cities are expanding rapidly by connecting to the Internet. These systems require openness in terms of their operation in the real world environment to achieve their functionality and benefits. Supporting openness creates many new research problems. All of our current composition techniques, analysis techniques, and tools need to be re-thought and re-developed to account for this type of openness, especially taking into account security and privacy. Human interaction is an integral part of openness. Consequently, openness must provide a correct balance between access to functionality, human interaction, and privacy and security. New research is needed to address specific dependencies of IoT apps on the environment, dealing with the realisms found in open environments, and solving the dependency problems that arise when many apps (for different purposes, for different people, etc.) are simultaneously executing.

**Including humans in the loop:** Many IoT applications will intimately involve humans, i.e., humans and things will operate synergistically. However, the dependencies between the human and the co-located and concurrently running apps must be addressed. For example, it is hypothesized that explicitly incorporating human-in-the-loop models for driving can improve safety, and using models of activities of daily living in home health care can improve medical conditions of the elderly and keep them safe via various assistive technologies. Although having humans in the loop has its advantage, modeling human behaviors is extremely challenging due to the complex physiological, psychological and behavioral aspects of human beings (humans do not follow electromechanical laws of nature). New research is necessary to raise human-in-the-loop control to a central principle in system and app design and to understand the complex dependencies between the apps and humans. In the formal methodology of feedback control there are several areas where a human model can be placed: outside the loop, inside the controller, inside the system model, inside a transducer, and at various levels in hierarchical control. In some IoT apps the human will play a role in all of these areas at the same time. Traditional control theory and supervisory control solutions are not adequate to deal with these issues. The newest challenge seems to be how to incorporate the human behavior as part of the system itself.

**Richer semantics of apps:** Assume IoT apps responsible for energy management (controlling thermostats, windows, and doors) and home health care (controlling lights, body nodes measuring heart rate and temperature, and sleep apnea machines) are running concurrently. Dependency analysis is required to avoid negative consequences. For example, the integrated system should not turn off medical appliances to save energy while they are being used as suggested by the home health care system. However, integrating multiple systems is very challenging as each individual system has its own assumptions, strategies to control the physical world, and semantics. For example, a home health care application may detect depression and decide to turn on all the lights. On the other hand, the energy management application may decide to turn off lights when no motion is detected. Detecting and resolving such dependency problems in real-time arising from the semantics of the apps is important for correctness of interacting IoT systems.

**Synchronization challenges for real-time analysis:** IoT systems include devices that communicate not only with the cloud but also with each other, often requiring real-time coordination and synchronization.

Consider a tele-health activity where a patient exercises at home and a physiotherapist remotely conducts an assessment of the patient's health during the prescribed exercises. The body health sensors (e.g., pulse, blood pressure, etc.) and room sensors (e.g., camera, microphone, etc.) capture data that together gives the therapist an understanding of the patient. All the captured sensory data will be very strongly correlated, taking essential data in synchronous manner about the patient that must be aligned exactly in real-time to provide the correct insights. This relatively simple scenario illustrates many challenges to existing infrastructure (OS, network, etc.) that make it difficult with current technology, where synchronizing fine-grain, complex information streams from multiple sources in real time was not in their original scope. For example, Networking Time Protocol (NTP) is being used for wide area network time resynchronization but only at the timing precision level of several milliseconds, which is not sufficient for IoT devices used in health care. New implementations, protocols, and standards are necessary to enable this rich class of applications.

**Cyber-Physical-Human Systems**

A cyber-physical-human (CPHS) system consists of a collection of computing devices communicating with one another and interacting with the physical world via sensors and actuators in a feedback loop with the possibility of human intervention, interaction and utilization. Such systems are central in many applications in the IoT, from smart buildings to medical devices to automobiles. The design and implementation of distributed cyber-physical-human systems requires an integrative approach from designing high-level abstractions with rigorous logical foundations, comprehensive human-centered design and the engineering solutions to realize these possibilities.

**Agile programming for CPHS systems:** Compared to traditional software systems, programming of cyber-physical systems has to address at least two challenges: it is a distributed system consisting of multiple interacting agents, and an agent is a hybrid system consisting of a discrete program interacting with a continuously evolving physical world. Note that designers of software for systems such as a modern aircraft are faced with similar challenges but in that context, the current system development process is highly specialized, proprietary, and very expensive. The cyber-physical systems in IoT applications demand an agile development process accessible to common programmers. A major challenge for improving programmer productivity and reducing development cost is to develop a programming (and system integration) environment that provides clearly defined primitives to design CPS applications from existing components. Developing a compiler for such a language is a challenging task due to the large gap between high-level programming abstractions and distributed heterogeneous execution platforms.

**Humans interacting with CPHS systems:** While humans need not be in direct control of the emerging CPS applications such as autonomous vehicles, they will definitely be supervisors, coaches, consumers, and collaborators. Because autonomous systems that perceive and reason about the real world are not always effective at representing uncertainties and interpreting ambiguous or conflicting information, applications must rely on effective human-autonomy interaction. This need for coordinated human-system interaction calls for research on developing principled approaches to leverage complementary strengths of humans and computers, and to develop new modes of communication between them to ensure synergistic operation. We also need a deeper understanding of the role of trust that the autonomous systems and humans have in each other.

An overarching challenge is the need for approaches that harmonize human interpretations of the cyber-physical world and machine interpretations of the human-physical world.  The messiness of networked appliances leaves digital traces of devices that do not mirror human understanding of physical boundaries (e.g. accidentally printing to the neighbor's wireless printer).  This messiness scales poorly as devices range from the 10s to the 100s and 1000s.   Automated systems attempting to recognize human behavior struggle to interpret "activities of daily living" for home-based healthcare and to reflect information boundaries based on collaborative activities in formal and informal settings.

**Control in the presence of adversaries:** The tight integration between distributed computational agents, the communication network, and the physical world with human actors amplifies the challenges in ensuring security and privacy. Given the safety-critical applications of CPS, it is necessary to ensure that the system remains safe even under attacks and continues to perform its functions, perhaps with reduced performance. Reasoning about control systems in the presence of new threat models will be increasingly important, especially as the systems become more complex with larger attack surfaces. Understanding attack models and developing attack-resilient state estimation and control algorithms is a new challenge for control theory. There is also need to explore the tradeoffs between protecting data for privacy and use of data to ensure safety.

## 4. Recommendations

IoT systems will create dramatic business opportunities and provide great benefits to individuals and society. For these systems to succeed, they must be secure, robust, and usable by humans.  Progress has been made on improving the security of existing systems but IoT systems require even higher quality and introduce new complexities. Existing methods to create high-integrity systems, such as aircraft control software, involve rigorous and costly certification processes.  Many IoT systems will have similar integrity requirements but will require much less time-consuming and expensive development processes. As we have outlined, significant advances in practices, tools, and development processes are required to achieve the greatest benefits from IoT.  Our recommendations are summarized below:

**Research Investments:**
- Invest in research to facilitate the construction, deployment, and automated analysis of **multi-component systems** with complex and dynamic dependences.  IoT systems by their nature will have dynamic membership and operate in unknown and unpredictable environments that include, by assumption, adversarial elements.
- **Going beyond formal methods research** (historically focusing on software and CPS) to create abstractions and formalisms for constructing and reasoning about systems with diverse and more difficult-to-characterize components such as human beings, machine learning models, data from crowds, etc.
- Support research that addresses the **core underlying scientific and engineering principles** dealing with large-scale issues, networking, security, privacy, impact of the physical on the cyber, real-time, and the other key questions raised in this document.

- Industry is application-focused and usually targets a single domain (health care, transportation, etc.). Support research that considers **architectures and solutions that transcend specific application domains**.
- Support research on the unique challenges and opportunities in **IoT security**, such as minimal operating systems to create IoT devices with smaller attack surfaces, new ways detect and prevent anomalous network traffic, and high-level policy languages for specifying permissible communication patterns.
- Invest in research in **cyber-human systems** that reflect human understanding and interaction with the physical world and (semi) autonomous systems.

Given the tremendous business opportunities for IoT technology, it is clear that investment, innovation, and wide adoption of this technology will occur at an unprecedented pace. As a result, the need for deep research investment in this area is time critical and major breakthroughs may have impact for decades to come. Companies always face challenging decisions about whether to build their own private infrastructure stack or leverage the benefits of open system ecosystems. As a result, it is essential that open standards based on robust and secure infrastructure emerge quickly to make it easier for companies to adopt open technology. It is essential that in addition to addressing the important research challenges outlined above, the community makes deep investments in collaboratively solving these many challenges.

**Community goals:**
- Create standards, incentives, frameworks, and infrastructures to **empower open environments and open interfaces** based on our current best understanding of secure and robust systems that result in a rich ecosystem of applications and devices. The importance of open interfaces and international standards is discussed at length in a recent ITU report entitled "Regulation and the Internet of Things"[9]. ITU-T has created a Global Standards Initiative on IoT specifically to promote global standards in this area.
- Create technology and policies that **bridge the wide gap between the unregulated Internet and regulated vertical domains**, such as medical devices. Finding ways to allow successful interoperation across these domains is a key challenge in creating an ecosystem for rapid but safe and secure IoT innovation.
- Create **shared testbeds, simulators, benchmarks, and code bases** to enable researchers to build real systems and conduct realistic (and repeatable) experiments to evaluate new ideas with reasonable investments of time and money.
- **Bring together interdisciplinary and transdisciplinary research engagements** to address the difficult adoption challenges inherent in IoT advances across many industries (e.g. healthcare, transportation, commerce).
- **Encourage public/private partnerships that allow deeper shared investment** in understanding and solving the challenges outlined. Companies have access to resources and information that are otherwise hard for academics to get. At the same time, solutions to the important cross-cutting challenges will create major new business opportunities. Examples of such partnerships include the $500 million N-CITE (National Computing and Insight Technologies Ecosystem) initiative proposed by SIA/SRC.

---

[9] *GSR discussion paper: Regulation and the Internet of Things*", ITU (International Telecommunication Union) / GSR (Global Symposium for Regulators) June 2015. http://www.itu.int/en/ITU-D/Conferences/GSR/Documents/GSR2015/Discussion_papers_and_Presentations/GSR_DiscussionPaper_IoT.pdf

The Internet of Things, like previous technology waves before it, is poised to create enormous new business opportunities and greatly improve our lives. These new systems will continue to challenge our best efforts to build them in the most secure and robust ways possible. We have outlined key research problems that we foresee playing an important role in this new IoT wave and presented recommendations about ways to respond to them. An active dialog between key constituents of the community, including government, industry, and academia will ensure that these challenges are met in a timely manner and that the Internet of Things has the greatest impact possible.

Acknowledgements: We thank Margaret Martonosi for her insightful comments and feedback.*For citation use*: Alur R., Berger E., Drobnis A. W., Fix L., Fu K., Hager G. D., . . . Zorn B. (2015). *System Computing Challenges in the Internet of Things*: A white paper prepared for the Computing Community Consortium committee of the Computing Research Association. http://cra.org/ccc/resources/ccc-led-whitepapers/13

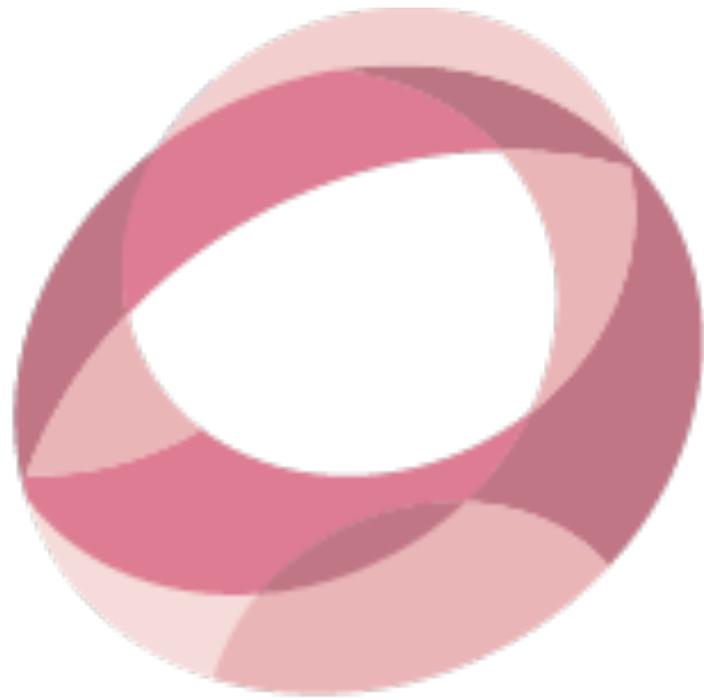

*This material is based upon work supported by the National Science Foundation under Grant No. (1136993). Any opinions, findings, and conclusions or recommendations expressed in this material are those of the author(s) and do not necessarily reflect the views of the National Science Foundation.*